\documentclass{revtex4}
\usepackage{color}
\usepackage{amsmath}
\usepackage{amssymb}
\usepackage{graphicx}
\usepackage{graphics}
\usepackage{bm}

\begin{document}

\title{Intersubband optical absorption in InSb stepped quantum wells. Effect
of spin sublevels crossing.}
\author{A Hern\'{a}ndez-Cabrera}
\email{ajhernan@ull.edu.es}
\author{P Aceituno}
\affiliation{Departamento de F\'{\i}sica, Universidad de La Laguna, La Laguna,
38206-Tenerife, Spain, and Instituto Universitario de Estudios Avanzados
(IUdEA) en F\'{\i}sica At\'{o}mica, Molecular y Fot\'{o}nica, Universidad de
La Laguna, La Laguna, 38206 Tenerife, Spain}

\begin{abstract}
We study linear and non- linear coefficients of the intersubband absorption
in InSb-based stepped quantum wells subjected to an in-plane magnetic field.
We consider also a transverse electric field to achieve near resonance
conditions. Taking into account the two deepest conduction levels and their
corresponding Zeeman spin splitting sublevels, we calculate dispersion
relations by means of an improved version of Kane model. Besides the known
anti-crossing between down and up spin split sublevels, we obtain an extra
spin level crossing for some determined parameters. This crossing clearly
modifies the absorption spectrum for transitions among the four sublevels
considered. We study a low electron density case, when only the first
deepest sublevel is occupied, and a high density case with only the highest
sublevel empty. We find a similar behavior of the absorption spectrum in
both cases.

\bigskip

\end{abstract}

\maketitle

\section{Introduction}

It is well-known that the key of spintronics is the breakdown of the
degenerate electronic levels by the spin splitting \cite{1}. This means that
spin up and spin down electronic states of any material must necessarily be
separated in energy. One way of achieving this splitting is to use
two-dimensional electron gas (2DEG) in quantum wells (QWs) . In
semiconductor QWs this effect is obtained spontaneously, without external
magnetic fields, as long as the confining potential is not symmetrical. Spin
splitting will increase due to the contribution of the Zeeman effect when an
external in-plane magnetic field is applied \cite{2,3}.

Some semiconductors are particularly suitable materials for spintronics. One
of them is the InSb with a large Land\'{e} factor (narrow gap), which causes
a big magnetic energy and the consequent Zeeman splitting.

Non symmetric heterostructures under in-plane magnetic field show non
parabolic dispersion relations. For each electronic level, non parabolic
spin split subbands with opposite spin are shifted by the magnetic field in
opposite directions of the momentum space. This behavior leads to the
presence of anticrossings between subbands for certain momentum values,
which are reflected in some peculiarities of the joint density of states and
in the excitation photoluminescence spectrum \cite{4,5,6}.

For particular structures with two close electronic levels, high enough
magnetic fields can cause a Zeeman splitting of each level bigger than the
interlevel energy distance. In this case, together with the momentum-space
displacements for different spin sublevels, we have different curvature of
quasi parabolas for different electronic levels leading to crossings between
the two distinct electronic levels with opposite spins. We should note that,
now, we are not talking about anticrossings but crossings, which has been
less studied \cite{7,8,9}.

To obtain close enough energy levels, with an intersubband energy distance
of the order of the spin splitting, we propose an InSb-based stepped QW.

The standard way to study the optical properties of a material is through
electronic transitions that occur after exposing the sample to a
perturbation, usually photoexcitation. From a theoretical point of view this
means a precise knowledge of the band structure and, hence, the dispersion
relations. Several methods were used to calculate the dispersion relations
in quantum structures. Among them, the Kane model together with the Transfer
Matrix Approximation (TMA), and first order boundary conditions \cite{10},
is particularly suitable for low-symmetry structures because of its
versatility. Moreover, this method allows us to add a wide range of
perturbations as transverse electric and in-plane magnetic fields. Another
effect we can add is abrupt barrier contribution for narrow gap structures.

Optical absorption is one of the most used experimental techniques to study
band structures. Absorption spectrum strongly depends on the electronic
concentration when heterostructure is selectively doped. Essentially,
electronic density is reflected in a variation of the frequency transition
between subbands.

The purpose of this work is the study of peculiarities of intersubband
optical absorption produced by the crossover of spin sublevels, including
electronic concentration effects.

\section{Theoretical framework}

\subsection{Eigenstates}

In the parabolic approximation, the one-electron Schr\"{o}dinger equation
for stepped QW can be written as \cite{6,11}: 
\begin{equation}
\left( \varepsilon ^{\mu }(\mathbf{p})+\frac{\widehat{p}_{z}^{2}}{2m_{\mu }}%
+U^{\mu }(z)+\widehat{W}^{\mu }(\mathbf{p})\right) \Psi ^{\mu }(\mathbf{p}%
,z)=E\Psi ^{\mu }(\mathbf{p},z)  \label{1}
\end{equation}%
for each 2D momentum $\mathbf{p}=\left( p_{x},p_{y}\right) $, where $\Psi
^{\mu }(\mathbf{p,}z\mathbf{)}$ and $E$, are the eigenfunctions and
eigenvalues, respectively. Superscript $\mu =b,w_{1},w_{2}$ means barrier or
wells (wide and narrow), respectively. The kinetic energy in the in-plane
direction, $\varepsilon ^{\mu }(\mathbf{p)}$ , includes the effective mass $%
m_{\mu }$. Band diagram for the stepped QW is shown in Fig. 1. 
\begin{figure}[h]
\centering\includegraphics[width=8cm]{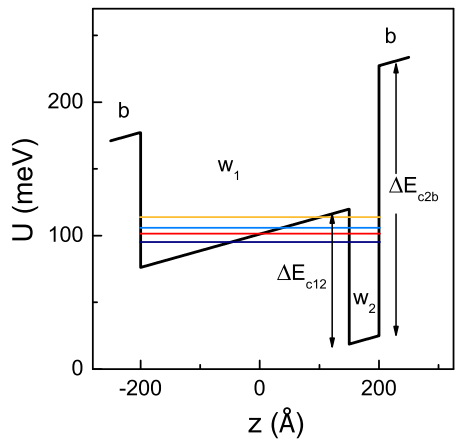}
\caption{Scheme of the stepped quantum well with the two deepest resonant
levels and their corresponding spin splitting. $b,w_{1},w_{2}$ means
barrier, wide well and narrow well. Band offsets are $\Delta E_{c12}$
between both wells, and $\Delta E_{c2b}$ between $w_{2}$ and barrier.}
\end{figure}
The potential energy $U^{\mu }(z)$ is $U^{\mu }(z)$ $\simeq U_{0}^{\mu
}+eF_{\bot }z$\ with$\ U_{0}^{b}=\Delta E_{c2b}$ in the barriers, $%
U_{0}^{w_{1}}=$ $\Delta E_{c12}$ in the wide well, and $U_{0}^{w_{2}}=0$ in
the narrow well. Here $F_{\bot }$\ is an uniform transverse electric field. $%
\Delta E_{c12}$ and $\Delta E_{c2b}$ \ are the band offsets for conduction
band between both wells, and between narrow well and barrier, respectively.

For not very strong magnetic fields, we describe the magnetic energy as $%
\widehat{W}^{\mu }=\overline{v}^{\mu }\left[ \mathbf{\hat{\sigma}}\times 
\mathbf{p}\right] _{z}+w_{H}^{\mu }\hat{\sigma}_{y}$, \noindent where $%
\mathbf{\hat{\sigma}}$ is the Pauli matrix, and $w_{H}^{\mu }=(\overline{g}%
^{\mu }/2)\mu _{B}H$ is the Zeeman splitting caused by the magnetic field.
Here $\overline{g}^{\mu }$ is the effective Land\'{e} factor, $\mu _{B}$ is
the Bohr magneton, and $H$ is the in-plane magnetic field. The
characteristic spin velocity for each layer is $\overline{v}^{\mu }=eF_{\bot
}\hbar /4m_{\mu }\varepsilon _{g}^{\mu }$, \noindent with $\varepsilon
_{g}^{\mu }$ the gap energy.

After some heavy algebra \cite{12}, we obtain the fundamental solutions of
Eq (1) for spin $\sigma =(\uparrow ,\downarrow )$: 
\begin{eqnarray}
\Psi ^{\mu \uparrow }(\mathbf{p,}z)=\left\{ \left[ a_{\mu \uparrow }Ai(\xi _{%
\mathbf{p}}^{\mu \uparrow })+b_{\mu \uparrow }Bi(\xi _{\mathbf{p}}^{\mu
\uparrow })\right]+\rho ^{\mu -}(\mathbf{p})\left[ a_{\mu \downarrow }Ai(\xi
_{\mathbf{p}}^{\mu \downarrow })+b_{\mu \downarrow }Bi(\xi _{\mathbf{p}%
}^{\mu \downarrow })\right] \right\}(1/\sqrt{2})  \notag \\
\Psi ^{\mu \downarrow }(\mathbf{p,}z)=\left\{ \rho ^{\mu +}(\mathbf{p})\left[
a_{\mu \uparrow }Ai(\xi _{\mathbf{p}}^{\mu \uparrow })+b_{\mu \uparrow
}Bi(\xi _{\mathbf{p}}^{\mu \uparrow })\right]+\left[ a_{\mu \downarrow
}Ai(\xi _{\mathbf{p}}^{\mu \downarrow })+b_{\mu \downarrow }Bi(\xi _{\mathbf{%
p}}^{\mu \downarrow })\right] \right\}(1/\sqrt{2}),  \label{2}
\end{eqnarray}

\noindent where $\rho ^{\mu \pm }(\mathbf{p})=(\overline{v}^{\mu }p_{\pm
}+w_{H}^{\mu })/iw^{\mu }(\mathbf{p})$, with $w^{\mu }(\mathbf{p})= [(%
\overline{v}^{\mu }p_{x}+w_{H}^{\mu })^{2}+( \overline{v}^{\mu
}p_{y})^{2}]^{1/2}$ and $p_{\pm }=p_{x}\pm ip_{y}$ . In the above equation $%
Ai(\xi _{\mathbf{p}}^{\mu \sigma })$ and$\ Bi(\xi _{\mathbf{p}}^{\mu \sigma
})$ are the Airy functions with arguments%
\begin{equation}
\xi _{\mathbf{p}}^{\mu \sigma }=\frac{z}{l_{\bot }^{\mu }}+\frac{\varepsilon
^{\mu \sigma }(\mathbf{p})-E+U_{0}^{\mu }}{\varepsilon _{\bot }^{\mu }}.
\label{3}
\end{equation}%
with the following auxiliary parameters: the length $l_{\bot }^{\mu }=\left(
\hbar ^{2}/2m_{\mu }eF_{\bot }\right) ^{1/3}$, and energies $\varepsilon
_{\bot }^{\mu }=\hbar ^{2}/[2m_{\mu }\left( l_{\bot }^{\mu }\right) ^{2}]$, $%
\ \varepsilon ^{\mu \uparrow }(\mathbf{p})=\varepsilon ^{\mu }(\mathbf{p}%
)+|w^{\mu }(\mathbf{p})|$, and $\varepsilon ^{\mu \downarrow }(\mathbf{p}%
)=\varepsilon ^{\mu }(\mathbf{p})-|w^{\mu }(\mathbf{p})|$.\noindent\ Lastly, 
$a_{\mu \sigma }$, $b_{\mu \sigma }$ are unknown coefficients that we will
obtain by means of the boundary conditions, including abrupt interface
parameter \cite{13} $\chi ^{\mu \nu }=(2eF_{\bot }\delta +\left\vert
U_{0}^{\mu }-U_{0}^{\nu }\right\vert )/2\varepsilon _{g}\approx \left\vert
U_{0}^{\mu }-U_{0}^{\nu }\right\vert /2\varepsilon _{g}$, \noindent where $%
\delta $ is the halfwidth of that interface \cite{12}.

The following step is to generate $4\times 4$ Wronskian-like transfer
matrices, $M^{\mu }(L_{i},E,\mathbf{p}),$ which \ involve contour conditions
at interface $L_{i}.$ To obtain electronic levels for each 2D momentum $%
\mathbf{p}=\left( p_{x},p_{y}\right) $ we introduce a modification of the
method used before \cite{12}. The total transfer matrix can be written as: 
\begin{eqnarray}
S\left(E,\mathbf{p}\right)=[M^{b}(L_{1},E,\mathbf{p})]^{-1}\cdot
M^{w_{1}}(L_{1},E,\mathbf{p})\cdot[M^{w_{1}}(L_{2},E,\mathbf{p})]^{-1}\cdot 
\notag \\
M^{w_{2}}(L_{2},E,\mathbf{p})\cdot[M^{w_{2}}(L_{3},E,\mathbf{p})]^{-1}\cdot
M^{b}(L_{3},E,\mathbf{p}).  \label{4}
\end{eqnarray}
We obtain the exact solution of the Hamiltonian from 
\begin{equation}
\Omega \left( E,\mathbf{p}\right) =S_{11}\left( E,\mathbf{p}\right) \cdot
S_{33}\left( E,\mathbf{p}\right) -S_{31}\left( E,\mathbf{p}\right) \cdot
S_{13}\left( E,\mathbf{p}\right) =0  \label{5}
\end{equation}

The four roots of $\Omega \left( E,\mathbf{p}\right) $ are the\ solutions of
Eq. (1), $E_{k\sigma }\left( \mathbf{p}\right) $, which correspond to the
two deepest\ coupled levels of the stepped QW ($k=1,2$), and their
respective spin down and spin up sublevels ($\sigma =\uparrow ,\downarrow $%
). For a wide range of $\mathbf{p}$\ values we obtain dispersion relations.
We refer to them as quasi-paraboloids because pure paraboloid shape is
broken at anticrossing points \cite{4}. \ 

After obtaining coefficients $a_{\mu \sigma },\ b_{\mu \sigma }$\ we proceed
to calculate wave functions for each energy sublevel ($k\sigma $) and
momentum $\mathbf{p}$ (Eq. 2). If we denote by \ $\Psi ^{k\mu \sigma } (%
\mathbf{p,}z)$ the wave function $\Psi ^{\mu \sigma }(\mathbf{p,}z\mathbf{)}$
for a particular level $k$, then 
\begin{eqnarray}
\Psi ^{k\sigma }(\mathbf{p,}z)=\sum\limits_{\mu }\Psi ^{k\mu \sigma}(\mathbf{%
p,}z)=\Theta (L1-z)\Psi ^{kb\sigma }(\mathbf{p,}z)+\Theta (z-L1)\Theta
(L2-z)\Psi ^{kw_{1}\sigma }(\mathbf{p,}z)+  \notag \\
+\Theta (z-L2)\Theta (L3-z)\Psi ^{kw_{2}\sigma }(\mathbf{p,}z)+ \Theta
(z-L3)\Psi ^{kb\sigma }(\mathbf{p,}z),  \label{6}
\end{eqnarray}%
where $\Theta(z)$ is the Heaviside function. Finally, we normalize wave
functions.

Next we will analyze the intersubband absorption coefficient. Because this
coefficient is related to the transitions between occupied and empty
sublevels, we have to include electron density effects, which determine
occupied sublevels. Note that, till now, we have the one-electron solution.
Actually, for doped systems hamiltonian should include Hartree and Fock
potential terms, and Schr\"{o}dinger equation has to be solved
selfconsistently together with the Poisson equation.

To simplify calculations we assume that, for sheet electron densities $%
n_{2D} $ of the order of $10^{11}$ cm$^{-2}$, electron-electron interaction
does not alter one-electron results substantially. Thus, we will neglect
Fock term. For the Hartree potential, instead of using the
momentum-dependent self-consistency (with the difficulties involved due to $%
\mathbf{p}$ dependence), we solve the above one-electron Hamiltonian but
taking in mind the noticeable shift in the frequency that Hartree potential
produces in the intersubband electronic transitions. This shift is because
the photoexcitation electric field produces the superposition of wave
functions of the subbands involved in transitions. As a result, the charge
density is no longer homogeneously distributed along the $z$ direction. The
charge redistribution induces a space charge field that overlaps with the
laser driving field and affects the interlevel distance. This process is
known as depolarization \cite{14,15,16,17},. We define the depolarization
for the transition $\left( k\sigma \right) \rightarrow \left( k^{\prime
}\sigma ^{\prime }\right) $ as 
\begin{equation}
\delta _{k^{\prime }\sigma ^{\prime }k\sigma }=\frac{8\pi e^{2}\left(
n_{k\sigma }-n_{k^{\prime }\sigma ^{\prime }}\right) }{\epsilon \left(
E_{k^{\prime }\sigma ^{\prime }k\sigma }\right) }\int_{-\infty }^{\infty }dz%
\left[ \int_{-\infty }^{z}dz^{\prime }\sum\limits_{\mathbf{p}}\Psi
^{k^{\prime }\sigma ^{\prime }}(\mathbf{0,}z^{\prime }\mathbf{)}\Psi
^{k\sigma }(\mathbf{0,}z^{\prime }\mathbf{)}\right] ^{2}.  \label{10}
\end{equation}%
where \ $E_{k^{\prime }\sigma ^{\prime }k\sigma }=E_{k^{\prime }\sigma
^{\prime }}\left( \mathbf{0}\right) -E_{k\sigma }\left( \mathbf{0}\right) $, 
$\epsilon $ is the permittivity, and $n_{k\sigma }$ is the electron density
of the $k\sigma $\ sublevel.\ \ Considering the depolarization shift, the
renormalized interlevel energy can be expressed as 
\begin{equation}
\widetilde{E}_{k^{\prime }\sigma ^{\prime }k\sigma }=E_{k^{\prime }\sigma
^{\prime }k\sigma }\left( 1+\delta _{k^{\prime }\sigma ^{\prime }k\sigma
}\right) ^{1/2}.  \label{11}
\end{equation}

\subsection{Intersubband infrared absorption}

To analyze the absorption coefficient we have adapted Ahn and Chuang
expressions \cite{18}, obtained with matrix density formalism for parabolic
dispersion relations. In our case we deal with non-parabolic dispersion
relation and momentum-dependent integrals must be done numerically because
the loss of symmetry in the $\mathbf{p}$-space. However, calculations are
simplified because intersubband transitions between the fundamental and
first excited subbands are induced only by light incident parallel to the
growth plane \cite{19}. To say, the polarization vector lies in the $z$ axis
direction. In our case, the linear absorption coefficient for the optical
transition between states $\left( k\sigma \right) $ and $\left( k^{\prime
}\sigma ^{\prime }\right) $, as a function of the incident light frequency
and for fixed transverse electric and in-plane magnetic fields, reads 
\begin{equation}
\alpha ^{(1)}(\omega )=\omega \sqrt{\frac{\mu }{\epsilon }}\frac{2}{V}%
\sum_{k\sigma ,k^{\prime }\sigma ^{\prime }}\sum_{\mathbf{p}}\left\vert
M_{k^{\prime }\sigma ^{\prime }k\sigma }\left( \mathbf{p}\right) \right\vert
^{2}\frac{\left[ f_{k\sigma }(\mathbf{p)}-f_{k^{\prime }\sigma ^{\prime }}(%
\mathbf{p)}\right] \ \Gamma }{\left( \widetilde{E}_{k^{\prime }\sigma
^{\prime }k\sigma }-\hbar \omega \right) ^{2}+\Gamma ^{2}},  \label{12}
\end{equation}%
where%
\begin{equation}
M_{k^{\prime }\sigma ^{\prime }k\sigma }\left( \mathbf{p}\right) =\left\vert
e\right\vert \left\langle k^{\prime }\sigma ^{\prime }\left\vert
z\right\vert k\sigma \right\rangle =\left\vert e\right\vert \int_{-\infty
}^{\infty }zdz\ \left[ \Psi ^{k^{\prime }\sigma ^{\prime }}(\mathbf{p,}z%
\mathbf{)}\right] ^{\ast }\ \Psi ^{k\sigma }(\mathbf{p,}z\mathbf{)}
\label{13}
\end{equation}%
are the dipole matrix elements, and%
\begin{equation}
f_{k\sigma }(\mathbf{p)=}\frac{1}{1+\exp \left[ \left( E_{k\sigma }\left( 
\mathbf{p}\right) -\varepsilon _{F}\right) /k_{B}T\right] }  \label{14}
\end{equation}%
is the well known Fermi-Dirac function for a Fermi energy $\varepsilon _{F}$%
. The energy broadening of the absorption peaks is $\Gamma =\hbar /\tau _{r}$
where $\tau _{r}$ is the intersubband relaxation time. For simplicity, we
take an unique value for all transitions. In Eq. (12) $\mu $ is the
permeability and $\epsilon $\ is the permittivity of the wells (we take the
same values for the two materials), $c$ is the light speed in the vacuum,
and $V$ is the volume of the sample.

The third order nonlinear optical absorption is given by 
\begin{eqnarray}
\alpha ^{(3)}(\omega ,I)=-\omega \sqrt{\frac{\mu }{\epsilon }}\frac{2}{V}
\left( \frac{I}{2\epsilon n_{r}c}\right)\sum_{k\sigma ,k^{\prime }\sigma
^{\prime }}\sum_{\mathbf{p}}\left\vert M_{k^{\prime }\sigma ^{\prime
}k\sigma }\left( \mathbf{p}\right) \right\vert ^{4}\frac{\left[ f_{k\sigma
}( \mathbf{p)}-f_{k^{\prime }\sigma ^{\prime }}(\mathbf{p)}\right] \ \Gamma 
}{ \left[ \left( \widetilde{E}_{k^{\prime }\sigma ^{\prime }k\sigma }-\hbar
\omega \right) ^{2}+\Gamma ^{2}\right] ^{2}}\times  \notag \\
\left\{ 4-\frac{\left\vert M_{k^{\prime }\sigma ^{\prime }k^{\prime }\sigma
^{\prime }}\left( \mathbf{p}\right) -M_{k\sigma k\sigma }\left( \mathbf{p}%
\right) \right\vert ^{2}}{\left\vert M_{k^{\prime }\sigma ^{\prime }k\sigma
}\left( \mathbf{p}\right) \right\vert ^{2}} \frac{\left[ \left( \widetilde{E}%
_{k^{\prime }\sigma ^{\prime }k\sigma }-\hbar \omega \right) ^{2}-\Gamma
^{2}+2\widetilde{E}_{k^{\prime }\sigma ^{\prime }k\sigma }\left( \widetilde{E%
}_{k^{\prime }\sigma ^{\prime }k\sigma }-\hbar \omega \right) \right] }{%
\left[ \left( \widetilde{E}_{k^{\prime }\sigma ^{\prime }k\sigma }\right)
^{2}+\Gamma ^{2}\right] }\right\} ,  \label{15}
\end{eqnarray}%
where $I$ is the optical power per unit area. The total absorption
coefficient is%
\begin{equation}
\alpha (\omega ,I)=\alpha ^{(1)}(\omega )+\alpha ^{(3)}(\omega ,I).
\label{16}
\end{equation}%
Intersubband optical transitions between sublevels with the same spin $%
\left( \sigma \rightarrow \sigma \right) $ are called spin conserving
transitions, while those that occur between different spin sublevels $\left(
\sigma \rightarrow \sigma ^{\prime }\right) $ are often called spin flip
transitions.

\section{Results and discussion}

The structure we use in the calculations consists of a stepped QW formed by
a $400$ \AA\ wide QW of $In_{0.9}Al_{0.1}Sb$ ($w_{1}$) which includes a $50$ 
\AA\ wide QW of InSb ($w_{2}$). The structure is enclosed by $%
In_{0.8}Al_{0.2}Sb$ barriers. Data for this structure are $\Delta
E_{c12}=101.2$ meV, $\Delta E_{c2b}=202.3$ meV (Fig. 1), $%
m_{w_{1}}=0.0246m_{e}$, $m_{w_{w}}=0.0142m_{e}$, and $m_{b}=0.0352m_{e}$,
where $m_{e}$ is the free electron rest mass \cite{20,21}.

For this stepped QW resonance between ground and first excited conduction
levels, in absence of magnetic field, is achieved around an electric field $%
F_{\bot }=12.5$ meV. In addition to this transverse electric field we apply
an in-plane magnetic field to get Zeeman splitting. We find the desired
sublevels crossing for $H=6$ T. For higher magnetic fields crossings will
occurs at bigger $\mathbf{p}$ values where first sublevel is empty. For
lower magnetic fields there are not spin crossing. Thus, we also use $H=4$ T
to compare with the former case.

We consider the structure is selectively doped. To analyze concentration
effects we use two density values, $n_{2D}=8.25\times 10^{10}$ cm$^{-2}$ and 
$n_{2D}=3.6\times 10^{11}$ cm$^{-2}$, corresponding to $\varepsilon _{F}=103$
meV and $\varepsilon _{F}=113$ meV, respectively.

\subsection{Dispersion relations}

First we look for eigenenergy values by means of Eq. (5). Energy sublevels
correspond to the roots of $\Omega \left( E,\mathbf{p}\right) $ for each $%
\mathbf{p}$ value. Looking for the roots over a wide range of momenta we
obtain the dispersion relations. Because of the difficulty presented by the
quasi-paraboloids to see clearly\ spin crossings, we draw 2D sections of
them, without loss of generality. Fig 2 shows $E_{k\sigma }$versus $%
p_{x}/p_{0}$ for $H=6$ T and $p_{y}=0$. We take $p_{0}=m_{w_{2}}\overline{v}%
^{w_{2}}$ as a normalization factor to get dimensionless momentum. Arrows
indicate position of anticrossings of the two spin orientations of each
energy level. Due to the verticality of the curves in the region where they
occur, deformation of the parabolas in the area can not be perceived. Since
these anticrossings are not the aim of the present work we focus on the $%
\mathbf{p}$ region where crossings occurs.

\begin{figure}[h]
\centering\includegraphics[width=8cm]{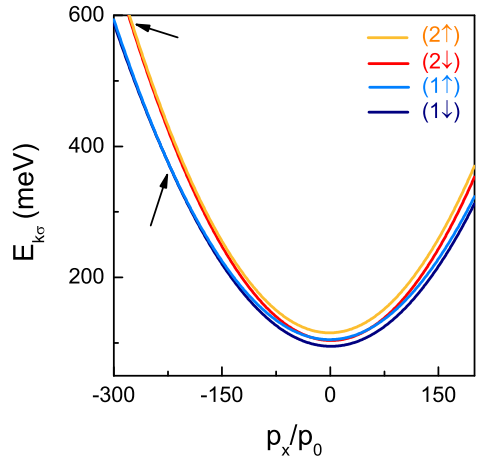}
\caption{Dispersion relations for $H=6$ T. Energy vs dimensionless $p_{x}$
for $p_{y}=0$. Arrows indicate anticrossing regions.}
\end{figure}

We present crossing region of the former figure in Fig 3(a). Looking at $%
p_{x}=0$, we can see the $\left( 1\uparrow \right) $ parabola is above the $%
\left( 2\downarrow \right) $ one. Since the curvature of the $\left(
2\uparrow \downarrow \right) $ parabolas is greater than that of the $\left(
1\uparrow \downarrow \right) $ parabolas and they shift in opposite
directions along $p_{x}$ axis for different spin values, the crossing of
spin sublevels $\left( 1\uparrow \right) $ and $\left( 2\downarrow \right) $%
\ is obvious for a pair of momentum values. These spin crossings happen
around $p_{x}/p_{0}=-28$ and $p_{x}/p_{0}=54.$ Beyond these momenta the
normal spin order is recovered. For an analogous region, Fig. 3(b) shows $%
E_{k\sigma }$versus $p_{y}/p_{0}$ for $p_{x}=0$. Now, spin crossings are
symmetrically placed at $p_{y}/p_{0}=\pm 39$.

\begin{figure}[h]
\includegraphics[width=8cm]{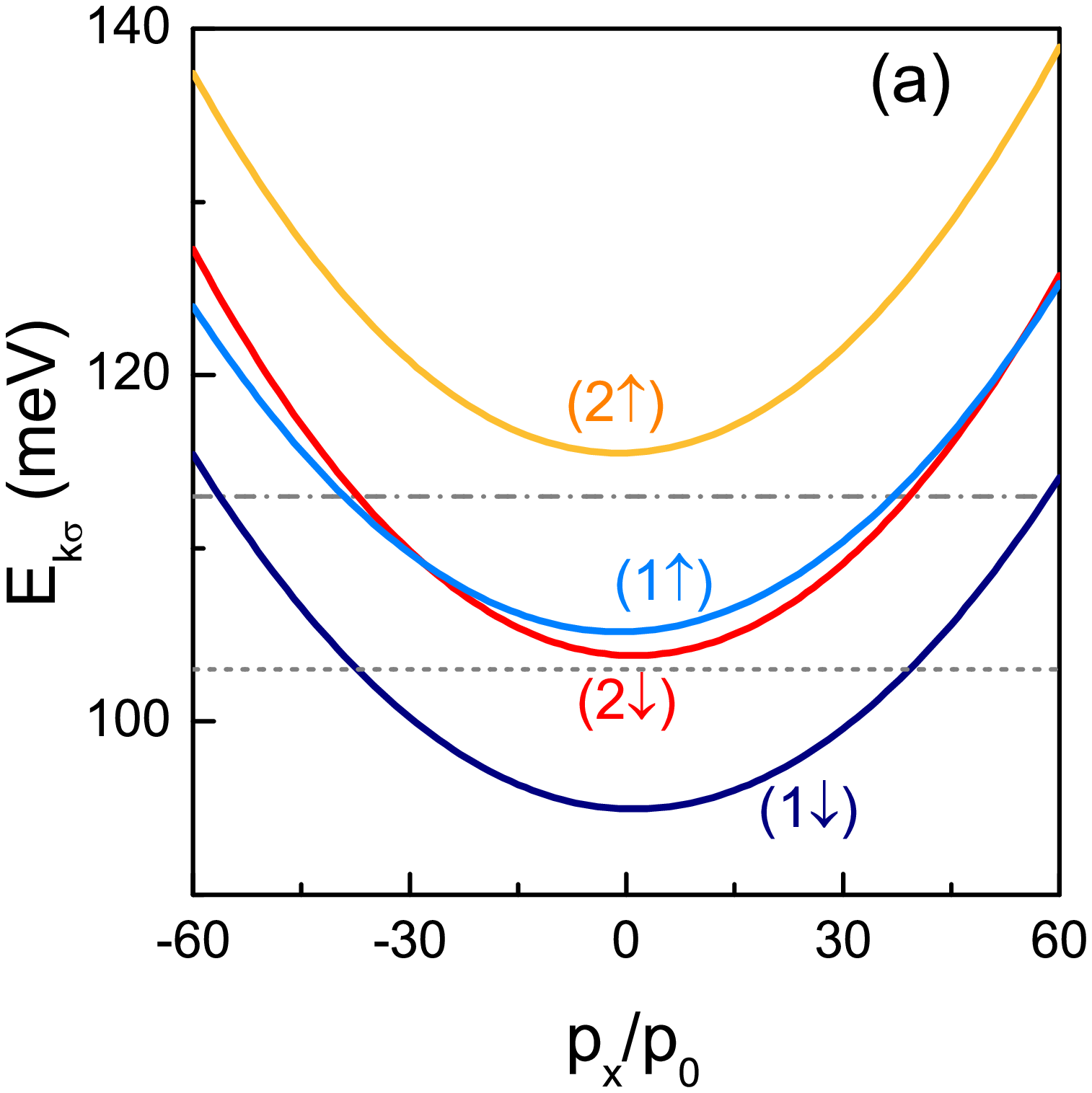} %
\includegraphics[width=8cm]{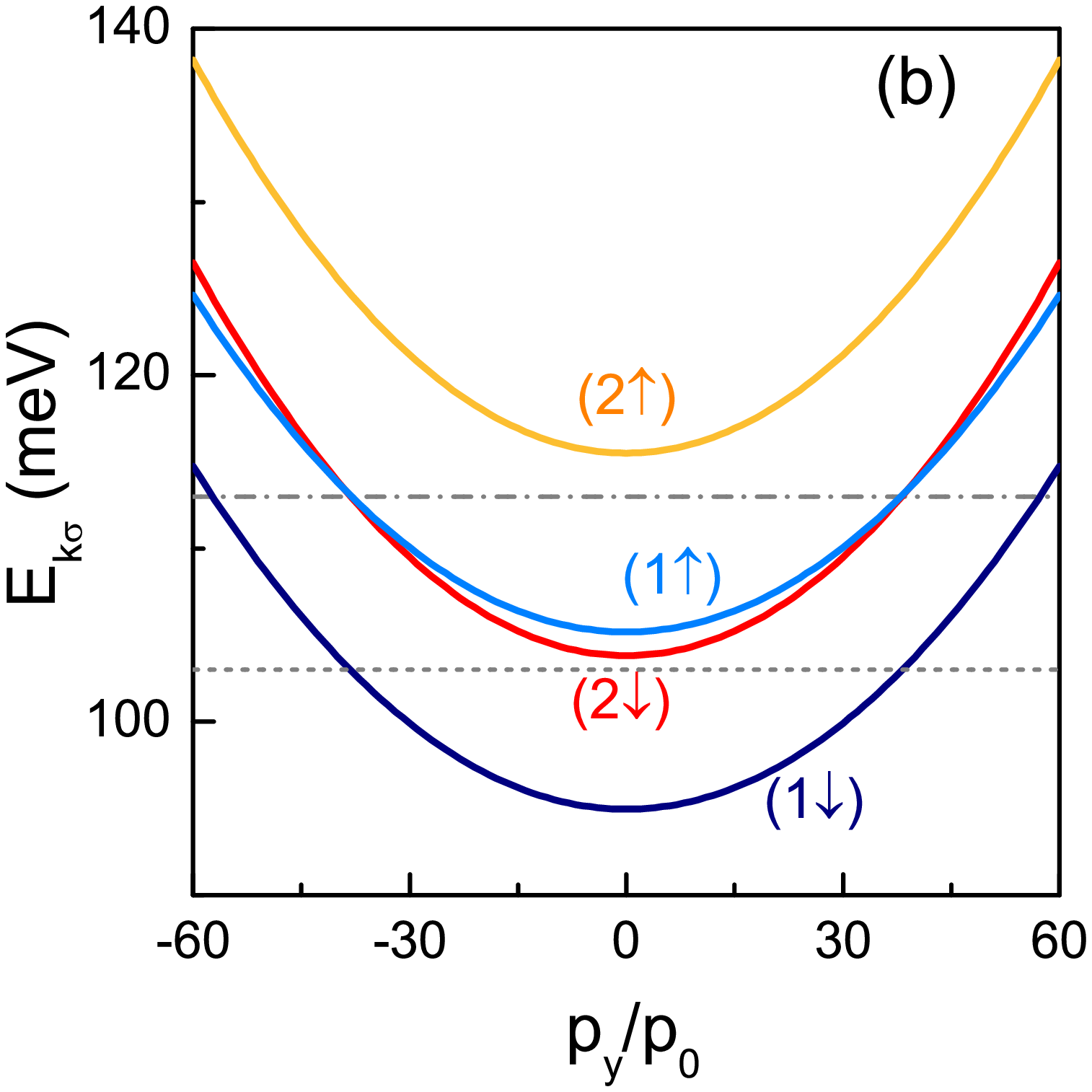}
\caption{Dispersion relations for $H=6$ T. (a) Energy vs dimensionless $%
p_{x} $ for $p_{y}=0$. Magnification of the central area of the previous
figure. (b) Energy vs dimensionless $p_{y}$ for $p_{x}=0$. Dotted line: $%
\protect\varepsilon _{F}=103$ meV, and dash-dotted line: $\protect%
\varepsilon _{F}=113$ meV.}
\end{figure}

In order to compare with the case where no spin crossings exist, we
represent in Fig. 4(a) and 4(b) relation dispersion for $H=4$ T. As can be
seen, in this case the $\left( 1\uparrow \right) $ parabola is under the $%
\left( 2\downarrow \right) $ one, following the usual behavior. Although
these parabolas approximate each other and it would seem that they are
crossing, the different curvature of parabolas leads to no crossing at all.

\begin{figure}[h]
\includegraphics[width=8cm]{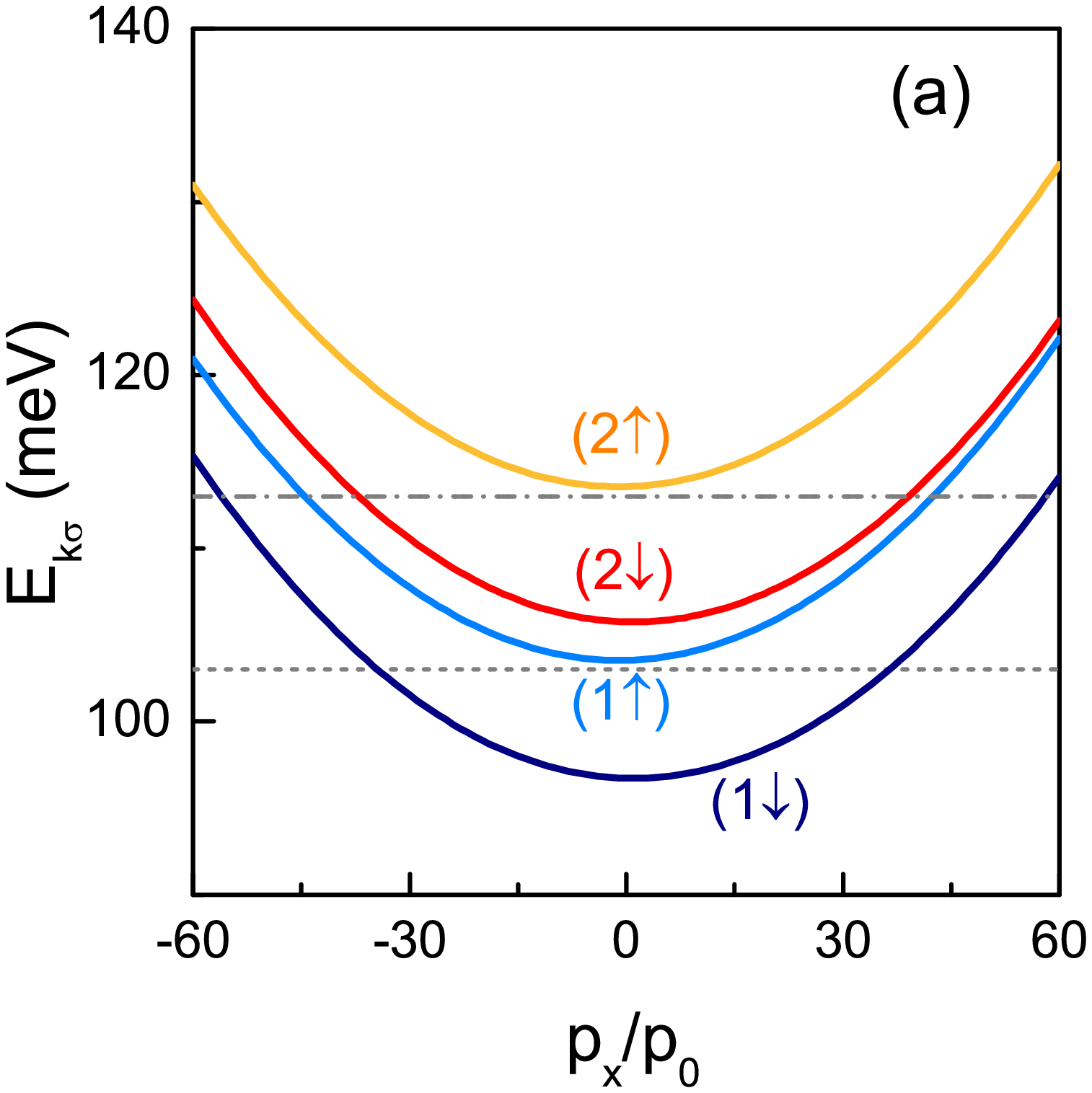} %
\includegraphics[width=8cm]{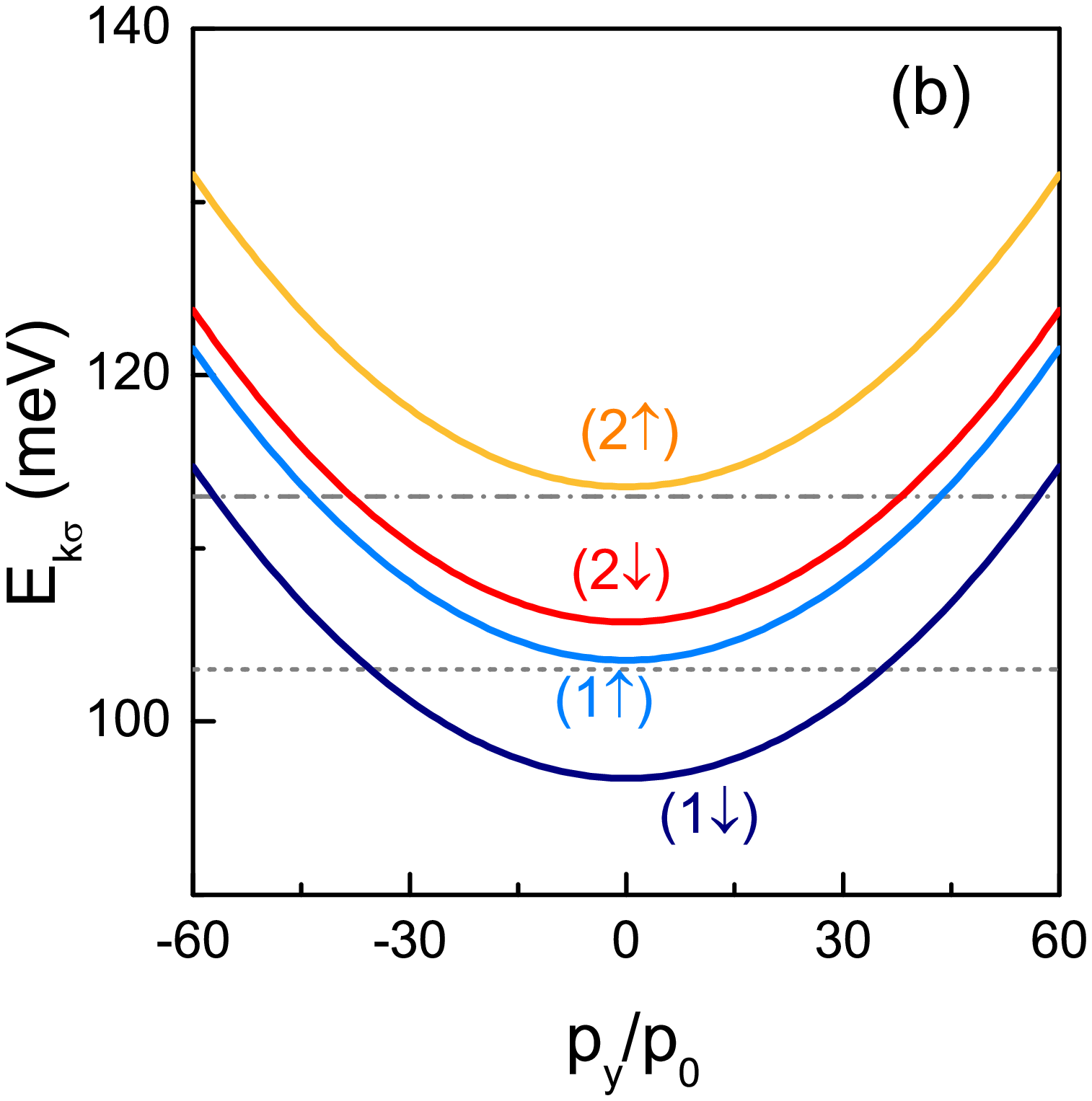}
\caption{Dispersion relations for $H=4$ T. (a) Energy vs dimensionless $%
p_{x} $ for $p_{y}=0$. (b) Energy vs dimensionless $p_{y}$ for $p_{x}=0$.
Dotted line: $\protect\varepsilon _{F}=103$ meV, and dash-dotted line: $%
\protect\varepsilon _{F}=113$ meV.}
\end{figure}

Once obtained dispersion relations we calculate and normalize wave functions 
$\Psi ^{k\sigma }(\mathbf{p,}z\mathbf{).}$ As an example Fig. 5 shows the
case for $H=6$ T and $\mathbf{p}=\mathbf{0}$. Due to the applied electric
field $F_{\perp }=12.5$ kV/cm, just after resonance, the first level is
mainly located in the left side of the wide well, whereas the second level
basically corresponds to the narrow well. In both cases sublevels $\sigma
=\uparrow $ are more confined in their corresponding wells, while sublevels $%
\sigma =\downarrow $ are more distributed between the two wells. This
behavior will affect the overlap of wave functions and thus, the
depolarization shift of transitions frequency.

\begin{figure}[h]
\centering\includegraphics[width=8cm]{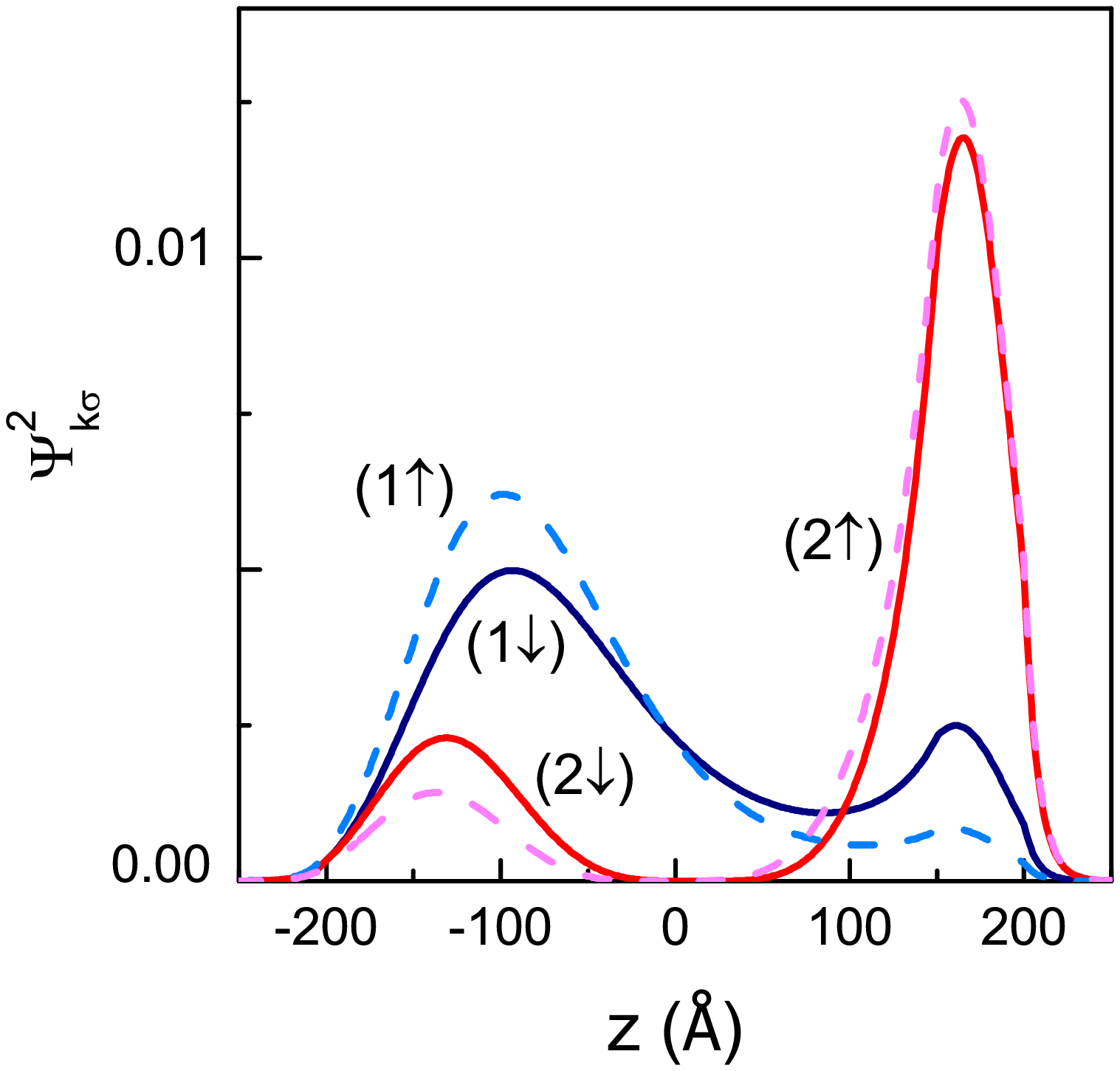}
\caption{Wave functions for $H=6$ T and $\mathbf{p}=\mathbf{0}$.}
\end{figure}

\subsection{Spin intersubband absorption}

Next, we calculate dipole matrix elements $M_{k^{\prime }\sigma ^{\prime
}k\sigma }\left( \mathbf{p}\right) $\ and, finally, the total absorption
coefficient taking into account both linear $\alpha ^{(1)}(\omega )$ and
non-linear $\alpha ^{(3)}(\omega ,I)$\ contributions. We take $\Gamma =1$
meV, $T=4.2$ K, and $I=1$ MW/cm$^{2}$.

First, we consider $n_{2D}=8.25\times 10^{10}$ cm$^{-2}$ corresponding to $%
\varepsilon _{F}=103$ meV. In this case only the deepest spin sublevel is
occupied, as can be seen in Figs. 3(a-b) and 4(a-b), for $H=6$ T and $H=4$
T, respectively. In these figures, Fermi level is represented by the dotted
line. Paraboloids corresponding to the three higher sublevels have energies
greater than Fermi energy and are empty. However, the deepest sublevel is
below Fermi energy in a certain momentum range. Electrons occupy this region
of the bottom of the paraboloid.

Figs. 6(a-b) show total absorption coefficient for $\varepsilon _{F}=103$
meV and when only the $\left( 1\downarrow \right) $ sublevel is occupied.
Fig. 6(a) presents the $H=6$ T case, where there is a spin crossing of
sublevels. Absorption spectrum shows three peaks corresponding to
transitions from this $\left( 1\downarrow \right) $ sublevel to the other
three empty sublevels $\left( 2\downarrow \right) $, $\left( 1\uparrow
\right) $ and $\left( 2\uparrow \right) $, respectively, in increasing order
of energy. The first is a spin-conserving transition while the other two are
spin flip transitions. Fig. 6(b) presents the other case, $H=4$ T, when
there is not such a crossing. Now, absorption spectrum displays two dominant
peaks, corresponding to transitions from $\left( 1\downarrow \right) $ to $%
\left( 1\uparrow \right) $ and $\left( 2\uparrow \right) $. The third peak,
for the transition $\left( 1\downarrow \right) $ to $\left( 2\downarrow
\right) $ is too small compared with the others and can not be noticed in
the total spectrum.

\begin{figure}[h]
\includegraphics[width=8cm]{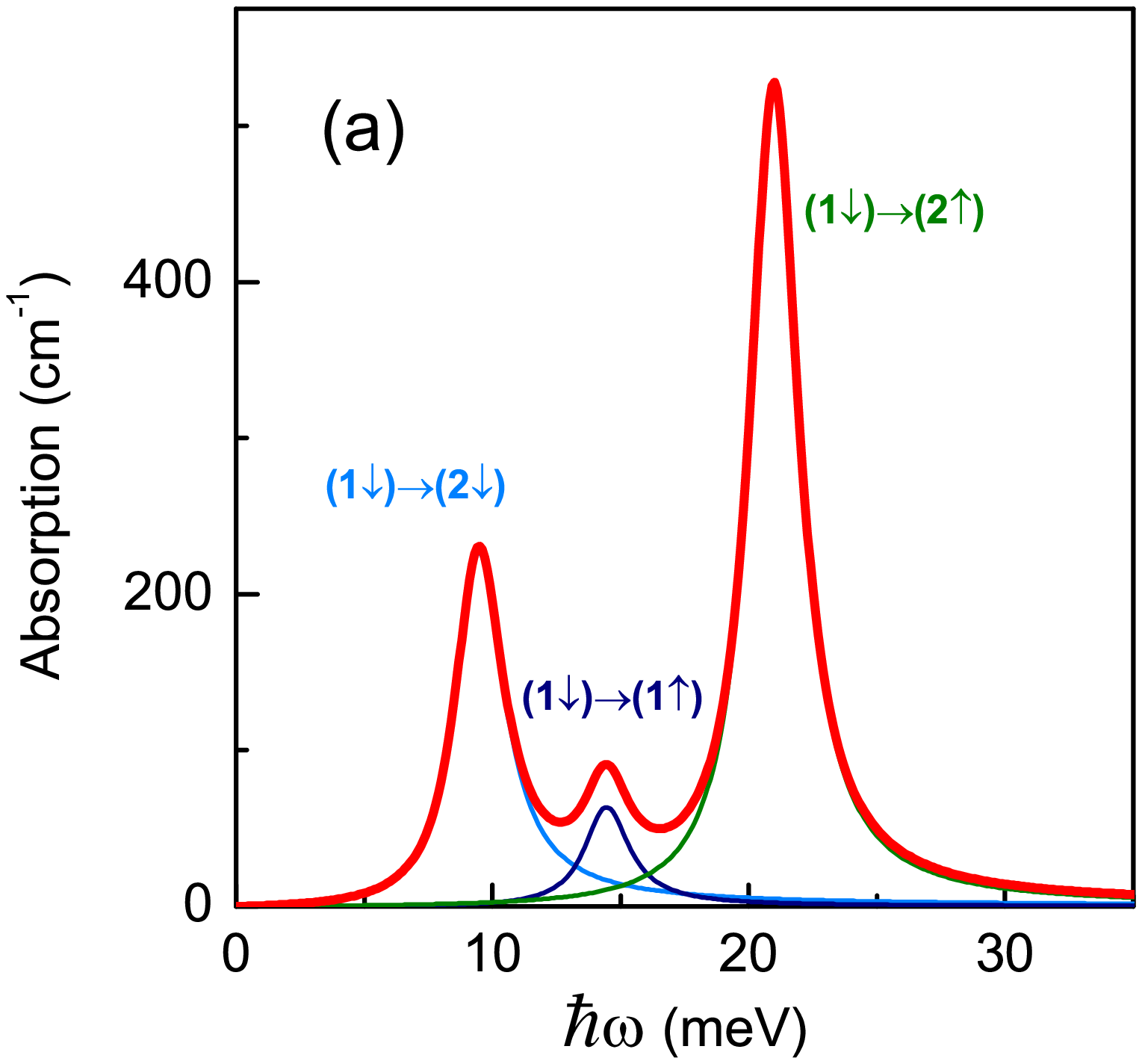} %
\includegraphics[width=8cm]{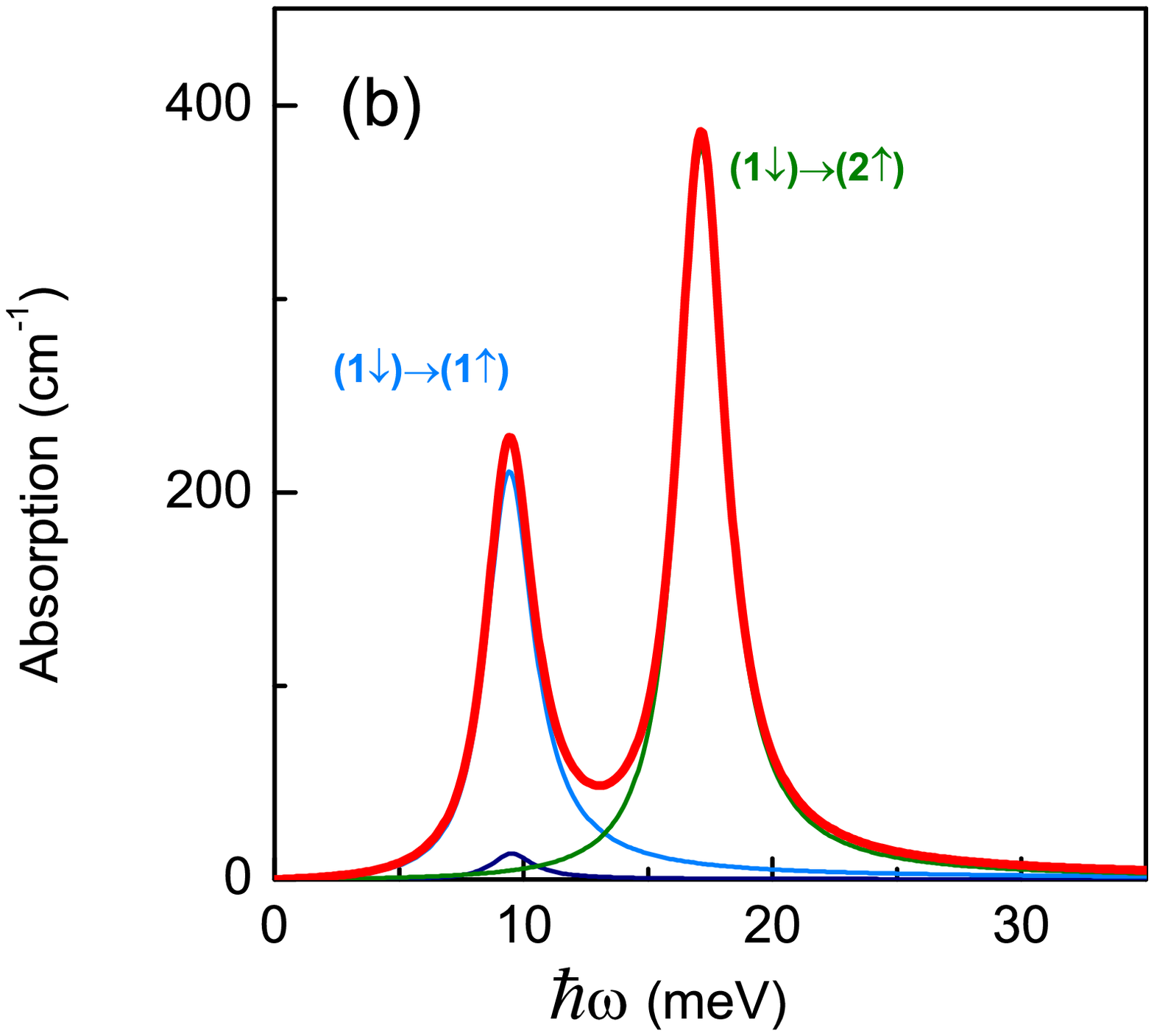}
\caption{Intersubband optical absorption for $\protect\varepsilon _{F}=103$
meV. (a) $H=6$ T. (b) $H=4$ T. }
\end{figure}

For $n_{2D}=3.6\times 10^{11}$ cm$^{-2}$, corresponding to $\varepsilon
_{F}=113$ meV, and due to levels proximity, we have three occupied
sublevels. This situation is also represented in dispersion relation figures
[Figs. 3(a-b) and 4(a-b)], where dash-dotted lines correspond to this Fermi
energy value. Unlike the previous case, now only paraboloid corresponding to
the highest sublevel has energy greater than Fermi energy and is the only
one completely empty. Nevertheless,\ the other three sublevels cut the Fermi
energy for some momentum values leading to the occupation of the bottom of
paraboloids with $E_{k\sigma }\left( \mathbf{p}\right) <\varepsilon _{F}$.

Figs. 7(a-b) present absorption coefficient for $\varepsilon _{F}=113$ meV.
Now, only sublevel $\left( 2\uparrow \right) $ is empty and transitions go
from the others sublevels to it. We have also considered the little regions
on left and right (in 2D figures but, actually, it is a thin ring in 3D
momentum space) of the dispersion relations where sublevels $\left(
1\uparrow \right) $ and $\left( 2\downarrow \right) $ lay over Fermi energy
level, while sublevel $\left( 1\downarrow \right) $ is under it. Thus, there
are two possible transitions between these sublevels in this $\mathbf{p}$%
-region, from the deepest one $\left( 1\downarrow \right) $ to the sublevels 
$\left( 1\uparrow \right) $ and $\left( 2\downarrow \right) $. Nevertheless,
results show that the peaks of these transitions are negligible compared
with the others and are not visible in the total spectrum. Fig. 7(a)
presents the "spin crossing case", for $H=6$ T. As in the former situation,
we find three peaks corresponding to transitions from spin sublevels $\left(
1\uparrow \right) $, $\left( 2\downarrow \right) $ and $\left( 1\downarrow
\right) $ to sublevel $\left( 2\uparrow \right) $, respectively, in the same
increasing order of energy. Fig. 7(b) shows the \ $H=4$ T case, where there
is not spin crossing. In a similar way as found before, we can appreciate
only two peaks, because the corresponding to spin-flip transition from $%
\left( 2\downarrow \right) $ sublevel to $\left( 2\uparrow \right) $ one is
smaller and, due to depolarization shift, almost coincides with the
transition from $\left( 1\uparrow \right) $ sublevel, resulting in a single
peak in the total spectrum. The other peak for the transition from $\left(
1\downarrow \right) $ to $\left( 2\uparrow \right) $ seems to be similar for
the two cases,\ since spin crossing sublevels are not involved.

\begin{figure}[h]
\includegraphics[width=8cm]{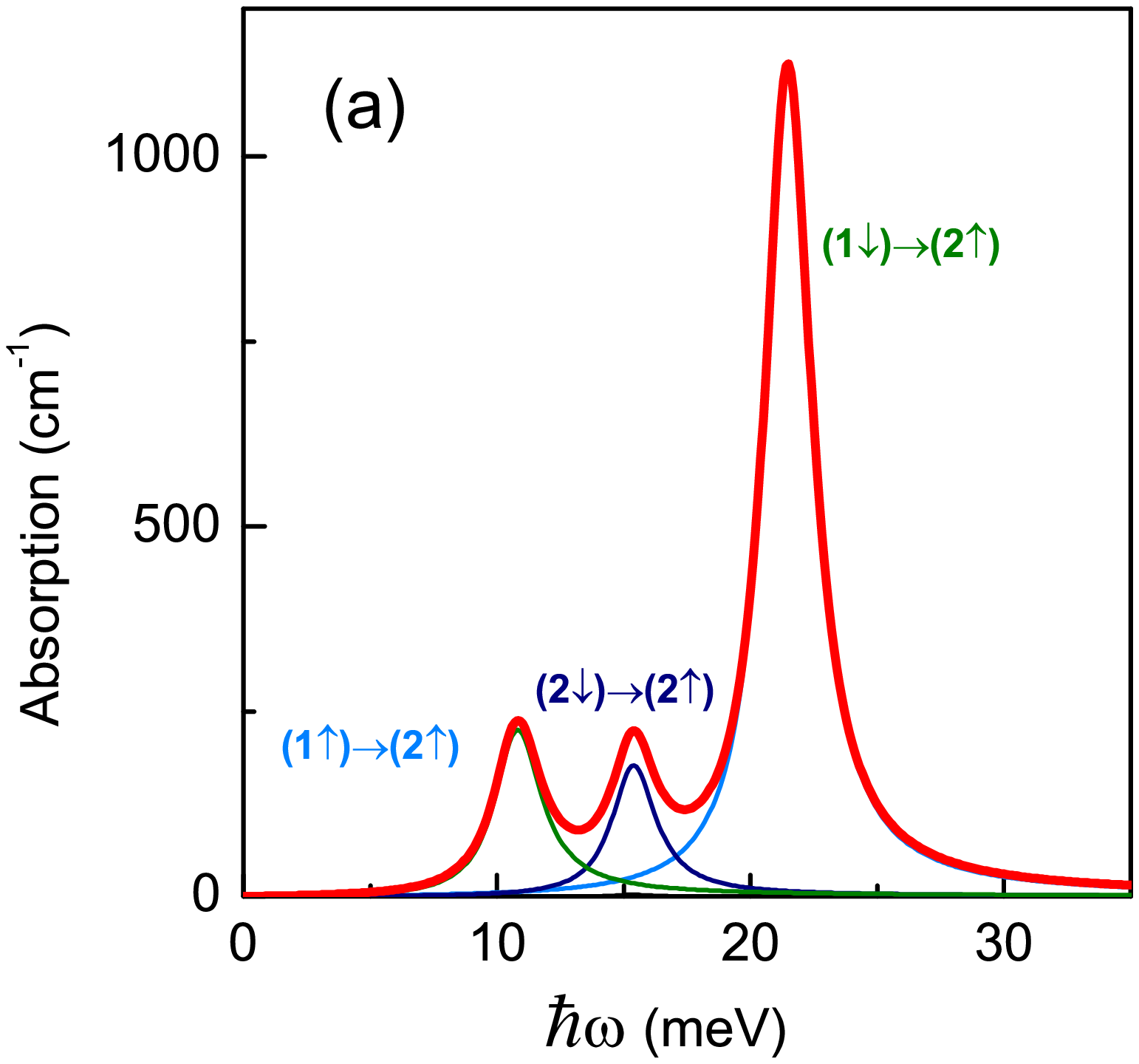}%
\includegraphics[width=8cm]{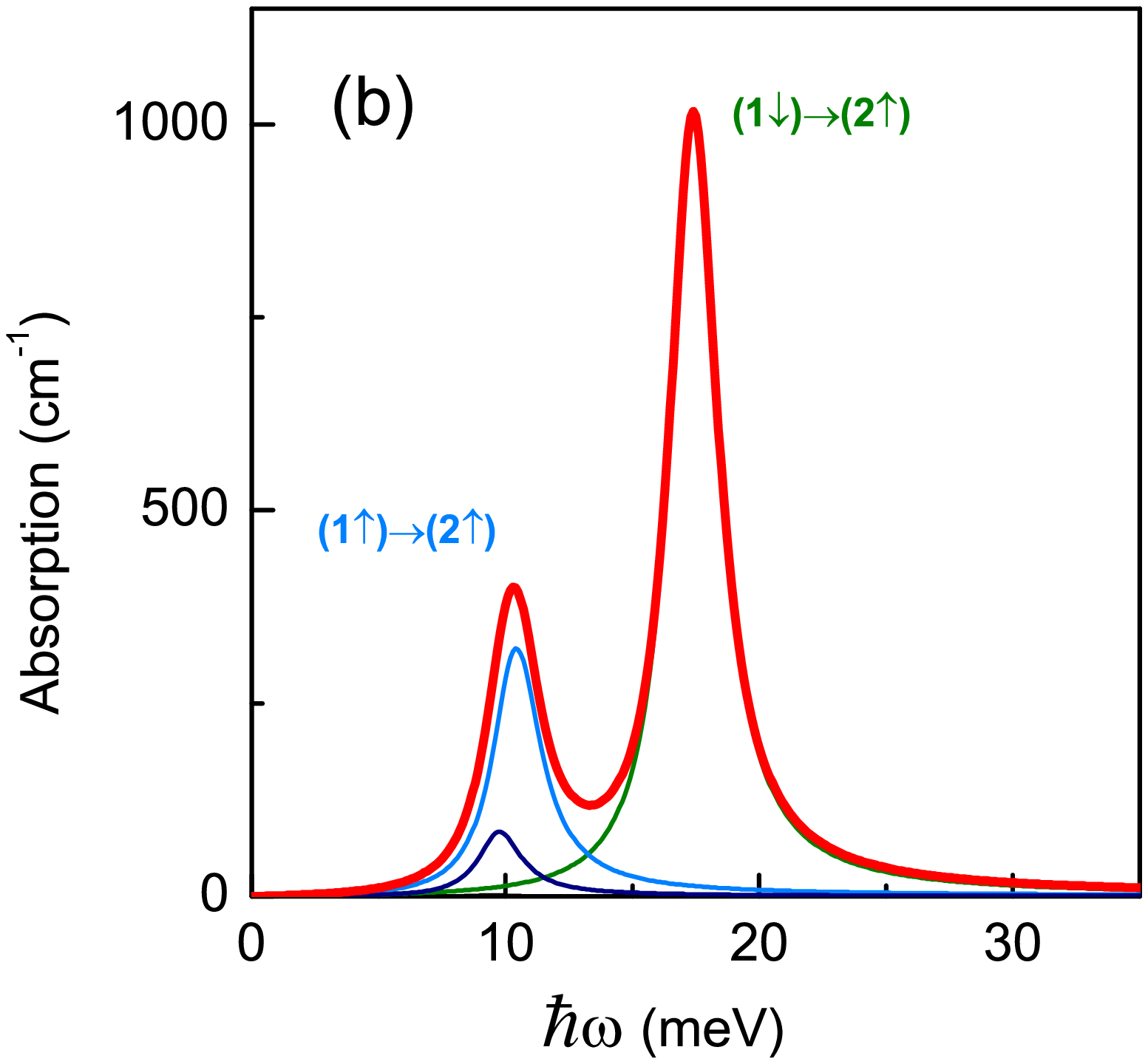}
\caption{Intersubband optical absorption for $\protect\varepsilon _{F}=113$
meV. (a) $H=6$ T. (b) $H=4$ T. }
\end{figure}

The behavior of the peaks, caused by the depolarization shift, can also be
seen for $\varepsilon _{F}=103$ meV in Fig. 6(b). The small peak
corresponding to spin-conserving transition $\left( 1\downarrow \right) $ to 
$\left( 2\downarrow \right) $, besides being hardly visible in the total
spectrum, almost coincides with the spin-flip transition $\left( 1\downarrow
\right) $ to $\left( 1\uparrow \right)$.

Considering the four cases under study we can see a general behavior both
for transitions from an unique occupied spin sublevel [$\left( 1\downarrow
\right) $ $\rightarrow $ $\left( 1\uparrow \right) $, $\left( 2\downarrow
\right) $, $\left( 2\uparrow \right) $], and for transitions up to an unique
empty sublevel [$\left( 1-\right) $, $\left( 1\uparrow \right) $ , $\left(
2-\right) \rightarrow $ $\left( 2\uparrow \right) $], as Figs. 6 and 7 show.
The transition between the lower and higher sublevels [$\left( 1\downarrow
\right) \rightarrow $ $\left( 2\uparrow \right) $] gives rise to the
dominant absorption peak in every case, specially for higher electronic
density. Because of the involved sublevels are not susceptible of crossing,
their behavior is similar for both magnetic fields used.

Sublevels $\left( 1\uparrow \right) $ and $\left( 2\downarrow \right) $ are
close together in energy. This would lead to the nearness of the transitions
involved: [$\left( 1\downarrow \right) $ $\rightarrow $ $\left( 1\uparrow
\right) $ , $\left( 2\downarrow \right) $] and [$\left( 1\uparrow \right) $
, $\left( 2\downarrow \right) \rightarrow $ $\left( 2\uparrow \right) $]
(for $\varepsilon _{F}=103$ meV and $\varepsilon _{F}=113$ meV,
respectively), and only one peak would be observable. However, when
considering the depolarization shift of transitions, absorption peaks
separate each other and become distinguishable in the spin crossing case.
Depolarization shift depends on wave functions overlapping. It can be proved
that overlap corresponding to spin sublevels of the same electronic level [$%
\left( 1\downarrow \right) $ $\rightarrow $ $\left( 1\uparrow \right) $] and
[$\left( 2\downarrow \right) $ $\rightarrow $ $\left( 2\uparrow \right) $]\
is very much greater than that of sublevels of different electronic levels [$%
\left( 1\downarrow \right) $ $\rightarrow $ $\left( 2\downarrow \right) $]
and [$\left( 1\uparrow \right) $ $\rightarrow $ $\left( 2\uparrow \right) $%
]. This fact is reflected in the different energy shifts of the
corresponding absorption peaks.

When spin crossing exists and for $\varepsilon _{F}=103$ meV [Fig. 6(a)],
absorption peaks follow the order in energy [$\left( 1\downarrow \right) $ $%
\rightarrow $ $\left( 2\downarrow \right) $], [$\left( 1\downarrow \right) $ 
$\rightarrow $ $\left( 1\uparrow \right) $]. By considering shift caused by
depolarization, displacement of second transition [$\left( 1\downarrow
\right) $ $\rightarrow $ $\left( 1\uparrow \right) $] is much greater than
that of the first one and both peaks separate each other and become clearly
visible in the total spectrum. A similar situation occurs for $\varepsilon
_{F}=113$ meV [Fig. 7(a)], where the order of transitions is [$\left(
1\uparrow \right) $ $\rightarrow $ $\left( 2\uparrow \right) $], [$\left(
2\downarrow \right) $ $\rightarrow $ $\left( 2\uparrow \right) $]. Again,
peak corresponding to second transition shifts more than that of the other
and both are discernible.

When there is not spin crossing the position of the transitions is inverted
[Figs. 6(b) and 7(b)]. In this case, the larger displacement of the first
peak makes it almost coincident with the second one, showing an unique peak
for both transitions. Results are in agreement with available experimental
data for the case where there is not spin sublevel crossing \cite{22} and
other theoretical works for stepped quantum wells \cite{23}.

\section{Conclusions}

In this work we calculate linear and non-linear intersubband optical
absorption coefficients in InSb-based stepped quantum wells, including
electron density effects through the displacements of transitions caused by
depolarization. We use a modified version of the Kane model together with
the TMA, which includes transverse electric field, in-plane magnetic field
and abrupt interfaces contribution to obtain the dispersion relations.
Considering Zeeman splitting of the electronic levels, we find spin sublevel
crossing or intersection of quasi-paraboloids for certain values of the
magnetic field. The existence of spin sublevel crossing essentially modifies
absorption spectrum: it changes the energy order of the electron transitions
which, together with the depolarization shift, provides an additional peak
in the structure of the absorption spectrum when compared with the standard
non crossing situation.

A similar theoretical analysis may be developed for the study of optical
properties in other structures under magnetic field. We hope that present
results will stimulate experimental efforts towards the study of spin
crossing peculiarities in nanostructures.


\begin{thebibliography}{99}
\bibitem{1} J. M. Kikkawa and D. D. Awschalom, "Resonant Spin Amplification
in n-Type GaAs", Phys. Rev. Lett. \textbf{80}(19), 4313-4316 (1998).

\bibitem{2} J. Xia, W. Ge, and K. Chang, \textit{Semiconductor Spintronics}
(World Scientific, 2012).

\bibitem{3} R. Winkler, \textit{Spin-Orbit Coupling Effects in
Two-Dimensional Electron and Hole Systems} (Springer-Verlag, Berlin, 2003).

\bibitem{4} A. Hern\'{a}ndez-Cabrera, P. Aceituno, and F. T. Vasko, "Level
anticrossing effect on electron properties of coupled quantum wells under an
in-plane magnetic field", Phys. Rev. B \textbf{60}(8), 5698-5704 (1999).

\bibitem{5} D. Huang and S. K. Lyo, "Photoluminescence spectra of n-doped
double quantum wells in a parallel magnetic field", Phys. Rev. B \textbf{59}%
, 7600(1999). S. K. Lyo, "Transport and level anticrossing in strongly
coupled double quantum wells with in-plane magnetic fields", Phys. Rev. B 
\textbf{50}(7), 4965-4968 (1994)

\bibitem{6} A. Hern\'{a}ndez-Cabrera, P. Aceituno, and F. T. Vasko,
"Electron energy spectrum and density of states for nonsymmetric
semiconductor heterostructures in an in-plane magnetic field", Phys. Rev. B 
\textbf{74}(3), 035330 (2006).

\bibitem{7} A. C. Graham, K. J. Thomas, M. Pepper, N. R. Cooper, M. Y.
Simmons, and D. A. Ritchie, "Interaction Effects at Crossings of
Spin-Polarized One-Dimensional Subbands", Phys. Rev. Lett. \textbf{91}(13),
136404 (2003).

\bibitem{8} K. F. Berggren, P. Jaksch, and I. Yakimenko, "Effects of
electron interactions at crossings of Zeeman-split subbands in quantum
wires", Phys. Rev. B \textbf{71}(11), 115303 (2005).

\bibitem{9} Y. V. Pershin, J. A. Nesteroff, and V. Privman, "Effect of
spin-orbit interaction and in-plane magnetic field on the conductance of a
quasi-one-dimensional system", Phys. Rev. B \textbf{69}(12), 121306(R)
(2004).

\bibitem{10} F. T. Vasko, "Spin splitting in the spectrum of two-dimensional
electrons due to the surface potential", JETP Lett. \textbf{30}(9), 541-544
(1979).

\bibitem{11} A. Hern\'{a}ndez-Cabrera, P. Aceituno, and F. T. Vasko,
"Quantum wells under an in-plane magnetic field: Effect of the composition
parameters on excited electron energy splitting", Journal of Luminescence 
\textbf{128}(5-6), 862-864 (2008).

\bibitem{12} A. Hern\'{a}ndez-Cabrera and P. Aceituno, "Abrupt barrier
contribution to electron spin splitting in asymmetric coupled double quantum
wells", Indian J. Phys. DOI 10.1007/s12648-014-0515-5 (2014).

\bibitem{13} F. T. Vasko and A. V. Kuznetsov, \textit{Electronic States and
Optical Transitions in Semiconductor Heterostructures} (Springer, New
York,1999).

\bibitem{14} M. Zaluzny, "Influence of the depolarization effect on the
nonlinear intersubband absorption spectra of quantum wells", Phys. Rev. {B 
\textbf{47}}(7), 3995-3998 (1993).

\bibitem{15} R. J. Warburton, C. Gauer, A. Wixforth, J. P. Kotthaus, B.
Brar, and H. Kroemer, "Intersubband resonances in InAs/AlSb quantum wells:
Selection rules, matrix elements, and the depolarization field", Phys. Rev.
B \textbf{53}(12), 7903-7910 (1996)

\bibitem{16} A. A. Batista, P. I. Tamborenea, B. Birnir, M. S. Sherwin, and
D. S. Citrin, "Nonlinear dynamics in far-infrared driven quantum-well
intersubband transitions" Phys. Rev. B \textbf{66}(19), 195325 (2002).

\bibitem{17} A. Hern\'{a}ndez-Cabrera and P. Aceituno, "Calculation of
intersubband absorption in doped graded quantum wells under intense
terahertz irradiation", Phys. Rev. B \textbf{78}(3), 035302 (2008).

\bibitem{18} D. Ahn and S. L. Chuang, "Calculation of linear and nonlinear
intersubband optical absorptions in a quantum well model with an applied
electric field", IEEE J. Quantum Electron. \textbf{23}(12), 2196-2204 (1987).

\bibitem{19} M. Virgilio and G. Grosso, "Optical transitions between valley
split subbands in biased Si quantum wells", Phys. Rev. B \textbf{75}(23),
235428 (2007).

\bibitem{20} D. W. Palmer, "The Semiconductors Information",
http://www.semiconductors.co.uk/

propiiiv5653.htm. Ioffe Physico-Technical Institute, "New Semiconductor
Materials. Characteristics and Properties",
http://www.ioffe.rssi.ru/SVA/NSM/Semicond/InSb/.

\bibitem{21} M. Edirisooriya, T. D. Mishima, C. K. Gaspe, K. Bottoms, R. J.
Hauenstein, M. B. Santos, "InSb quantum-well structures for electronic
device applications", J Crystal Growth \textbf{311}(7), 1972-1975 (2009).

\bibitem{22} M. B. Santos, S. D. Lowe, T. D. Mishima, R. E. Doezema, L. C.
Tung, and Y.-J. Wang, "Intersubband Absorption by Electrons in InSb Quantum
Wells with an In-Plane Magnetic Field", AIP Conf. Proc. \textbf{1399}, 133
(2011).

\bibitem{23} F T Vasko and G Y Kis, "The effect of a longitudinal magnetic
field on electronic intersubband transitions in asymmetric
heterostructures", Semiconductors \textbf{31}(9), 961-965 (1997).
\end{thebibliography}
\end{document}